\documentclass[twocolumn,english,aps,pra,showpacs,superscriptaddress]{revtex4}
\usepackage{pslatex}
\usepackage[T1]{fontenc}
\usepackage[latin1]{inputenc}
\usepackage{amssymb}

\makeatletter

\providecommand{\LyX}{L\kern-.1667em\lower.25em\hbox{Y}\kern-.125emX\@}
\usepackage{babel}
\usepackage{graphicx}
\makeatother
\input{epsf}
\begin{document}

\title{Pseudo-potential treatment
of two aligned dipoles under external harmonic confinement}

\author{K. Kanjilal}
\affiliation{Department of Physics and Astronomy, Washington State
University, Pullman, WA 99164-2814}
\author{John L. Bohn}
\affiliation{JILA, NIST and Department of Physics, University of 
Colorado, Boulder, CO 80309-0440}
\author{D. Blume}
\affiliation{Department of Physics and Astronomy, Washington State
University, Pullman, WA 99164-2814}
\affiliation{INFM-BEC, Dipartimento di Fisica, 
Universit\`a di Trento, Sommarive 4, I-38050 Povo, Italy}

\begin{abstract}
Dipolar Bose and Fermi gases, which
are currently being studied extensively 
experimentally and theoretically,  interact through anisotropic, long-range 
potentials. 
Here, we replace the long-range 
potential by a zero-range pseudo-potential that simplifies the
theoretical treatment of two dipolar particles in a harmonic trap.
Our zero-range pseudo-potential description reproduces
the energy spectrum of two dipoles 
interacting through a shape-dependent potential under external confinement
very well, provided that sufficiently many partial waves are included,
and readily leads to a classification scheme of the energy spectrum in
terms of approximate angular momentum quantum numbers. 
The results may be directly relevant to the physics of dipolar gases
loaded into optical lattices.
\end{abstract}

%34.10.+x General theories and models of atomic and molecular collisions and interactions (including statistical theories, transition state, stochastic and trajectory models, etc.)
%34.50.-s Scattering of atoms and molecules
%03.75.-b Matter waves (for atom interferometry techniques, see 39.20.+q-in atomic and molecular physics)
%\draft

\maketitle

\section{Introduction}
Many-body systems with dipolar interactions have attracted a lot of attention
recently. 
Unlike the properties of ultracold atomic alkali vapors,
which can be described to a very good approximation
by a single scattering quantity (the 
$s$-wave scattering length), those of dipolar 
gases additionally depend on the dipole moment.
This dipole moment can be magnetic,
as in the case of atomic Cr~\cite{grie05,stuh05}, or 
electric, as in the case of heteronuclear
molecules such as OH~\cite{meer05,boch04}, KRb~\cite{wang04a}
or RbCs~\cite{kerm04}.
Furthermore, dipolar interactions are long-ranged and anisotropic,
giving rise to a host of novel
many-body effects in confined dipolar gases such as
roton-like features~\cite{dell03,sant03,rone06a} and rich stability 
diagrams~\cite{sant00,yi00,gora00,mart01,yi01,gora02,rone06,bort06}.
The physics of dipolar gases loaded into optical
lattices promises to be particularly rich.
For example, this setup constitutes the starting point for a 
range of quantum
computing schemes~\cite{bren99,jaks00,demi02,bren02}.
Additionally, a variety of novel quantum phases have already been 
predicted to arise~\cite{gora02a,dams03,barn06,mich06}. 
Currently, a number of experimental groups are 
working towards loading dipolar gases into optical lattices.

This paper investigates the physics of doubly-occupied 
optical lattice sites in the regime where the tunneling between neighboring 
sites and the interactions with dipoles located in other lattice sites
can be neglected.
In this case, the problem reduces to 
treating the interactions between two dipoles in a single lattice site.
Assuming that the lattice potential can be approximated by a harmonic
potential, the center of mass motion separates
and the problem reduces to solving the Schr\"odinger 
equation for the relative distance vector
$\vec{r}$ between the two dipoles.
The interaction between the two aligned dipoles is angle-dependent 
and falls off as
$1/r^3$ at large interparticle distances. 
In this work, we replace the shape-dependent interaction potential by
an angle-dependent zero-range pseudo-potential, which is designed to
reproduce the scattering properties of the full shape-dependent
interaction potential, and derive
an implicit eigenequation
for two interacting identical bosonic dipoles and 
two interacting identical fermionic dipoles 
analytically.

Replacing the full
interaction potential or
a shape-dependent pseudo-potential by a
zero-range 
pseudo-potential~\cite{ferm34,huan57,busc98,blum02,bold02,kanj04,stoc04}
often allows for an analytical
description of ultracold two-body systems in terms of a few
key physical quantities. 
Here we show 
that the eigenequation for appropriately chosen
zero-range pseudo-potentials reproduces the energy spectrum of two dipoles
under harmonic confinement interacting through a shape-dependent
model potential; that
the applied zero-range treatment readily leads to
an approximate classification scheme of the energy spectrum in terms of
angular momentum quantum numbers; and
that the proposed pseudo-potential treatment breaks down
when the characteristic length of the dipolar interaction becomes
comparable to the characteristic length of the external confinement.
The detailed understanding of two interacting dipoles
obtained in this paper will guide optical lattice experiments
and the search for novel many-body effects. 

Section~\ref{sec_pp} introduces the Hamiltonian under study
and discusses the anisotropic zero-range pseudo-potential
that is used to describe the scattering between two interacting dipoles.
In Sec.~\ref{sec_ho}, we derive an implicit eigen equation for two
dipoles under external spherical harmonic confinement
interacting through the zero-range pseudo-potential 
and show that the resulting eigenenergies agree well with those
obtained for a shape-dependent model potential. Finally,
Sec.~\ref{sec_conclusion} concludes.

\section{System under study and anisotropic pseudo-potential}
\label{sec_pp}
Within the mean-field Gross-Pitaevskii 
formalism, the interaction between two
identical bosonic dipoles, aligned along the space-fixed
$\hat{z}$-axis by an external field, has been successfully 
modeled by the pseudo-potential $V_{pp}(\vec{r})$~\cite{yi00},
\begin{eqnarray}
\label{eq_dipole}
V_{pp}(\vec{r})= 
\frac{2 \pi \hbar^2}{\mu} a_{00} \delta(\vec{r})+
d^2 \frac{1-3 \cos^2 \theta}{r^3}.
\end{eqnarray}
Here, $\mu$ denotes the reduced mass of the two-dipole
system, $d$ the dipole moment,
and $\theta$ the angle
between $\hat{z}$ and the relative distance vector $\vec{r}$.
The $s$-wave scattering length $a_{00}$ depends on both the short- and
long-range parts of the true interaction potential.
The second term on the right hand side of Eq.~(\ref{eq_dipole})
couples angular momentum states with $l=l'$ ($l >0$) and 
$|l - l'| = 2$ (any $l,l'$). 
For identical fermions, $s$-wave scattering is absent
and the interaction 
is described, assuming the long-range dipole-dipole
interaction is dominant, by the second term on the right hand side of
Eq.~(\ref{eq_dipole}). 

Our goal in this paper is to determine 
the eigenequation of two identical bosonic dipoles
and two identical fermionic dipoles under
external spherically harmonic confinement with
angular trapping frequency $\omega$ analytically. 
The Schr\"odinger equation for the  
relative position vector $\vec{r}$
reads
\begin{eqnarray}
\label{eq_se}
[H_0 + V_{int}(\vec{r}) ] \psi(\vec{r}) = E \psi (\vec{r}),
\end{eqnarray}
where the Hamiltonian $H_0$ of the non-interacting 
harmonic oscillator is given by
\begin{eqnarray}
\label{eq_ham}
H_0 = -\frac{\hbar^2}{2 \mu} \nabla^2 _{\vec{r}} 
+\frac{1}{2}  \mu \omega^2 r^2.
\end{eqnarray}
In Eq.~(\ref{eq_se}), $V_{int}(\vec{r})$ denotes
the interaction potential.
The pseudo-potential $V_{pp}(\vec{r})$ cannot be used directly in
Eq.~(\ref{eq_se}) since both
parts of the pseudo-potential lead to divergencies.
The divergence of the $\delta$-function potential arises from
the singular $1/r$ behavior at small $r$ of the spherical Neumann
function $n_0(r)$, and can be cured by introducing the regularization operator
$\frac{\partial}{\partial r} r$~\cite{huan57}.
Curing the divergence of the long-ranged
$1/r^3$ term of $V_{pp}$
is more involved, since it couples an infinite number of angular
momentum states, each of which gives rise to a singularity in 
the $r \rightarrow 0$ limit. The nature of 
each of these singularities depends
on the quantum numbers $l$ and $l'$ coupled by the pseudo-potential,
and hence has to be cured separately for each $l$ and $l'$ combination.

In this work, we follow Derevianko~\cite{dere03,dere05} 
and cure the divergencies
by replacing $V_{pp}(\vec{r})$ with
a regularized zero-range potential $V_{pp,reg}(\vec{r})$,
which contains {\em{infinitely}} many terms,
\begin{eqnarray}
\label{eq_ppreg}
V_{pp,reg}(\vec{r}) = 
\sum_{ll'} V_{ll'}(\vec{r}).
\end{eqnarray}
The sum in Eq.~(\ref{eq_ppreg}) runs over $l$ and $l'$ even for
identical bosons, and over $l$ and $l'$ odd for identical
fermions. 
For $l \ne l'$, $V_{ll'}$ and $V_{l'l}$ are different and both
terms have to be included in the sum.
In Sec.~\ref{sec_ho}, 
we apply the pseudo-potential to systems under spherically symmetric
external confinement. For these systems, the projection quantum number
$m$ is a good quantum number, i.e., the energy spectrum for two interacting 
dipoles under spherically symmetric
confinement can be solved separately for each allowed $m$
value.
Consequently, a separate  pseudo-potential can be constructed for
each $m$ value.
In the following, we restrict ourselves to systems with vanishing
projection quantum number $m$;
the generalization 
of the pseudo-potential to general $m$ is discussed at the end 
of this section.
The $V_{ll'}$ are defined through their action
on an arbitrary $\vec{r}$-dependent function 
$\Phi(\vec{r})$~\cite{dere03,dere05},
\begin{eqnarray}
\label{eq_llprime}
V_{ll'}(\vec{r}) \Phi(\vec{r}) =
g_{ll'}
\frac{\delta(r)}{r^{l'+2}} 
Y_{l'0}(\theta,\phi) \times \nonumber \\
\left[
\frac{\partial^{2l+1}}{\partial r^{2l+1}} r^{l+1} 
\int Y_{l0}(\theta,\phi) \Phi(\vec{r}) d \Omega
\right]_{r \rightarrow 0}
\end{eqnarray}
with
\begin{eqnarray}
\label{eq_ppregstrength}
g_{ll'} = \frac{\hbar^2}{2 \mu} \frac{a_{ll'}}{k^{l+l'}} 
\frac{(2l+1)!! (2l' +1)!!}{(2l+1)!},
\end{eqnarray}
where
$k$ denotes the relative wave vector, $k=\sqrt{2 \mu E/\hbar^2}$,
and the
$a_{ll'}$ generalized scattering lengths.
Since we are restricting ourselves to
$m=0$, the $V_{ll'}$ are written in
terms of the spherical harmonics $Y_{lm}$ with $m=0$. 
When applying the above pseudo-potential 
we treat a
large number of terms in Eq.~(\ref{eq_ppreg}), 
and do not terminate the sum after the first
three terms 
as done in Refs.~\cite{dere03,dere05,yi04}.
We note that the non-Hermiticity of $V_{pp,reg}$
does not lead to problems when determining the
energy spectrum; however, great care has to be taken when
calculating, e.g., structural expectation values~\cite{reic06}.

To 
understand the functional form of the zero-range pseudo-potential
defined in Eqs.~(\ref{eq_ppreg}) through (\ref{eq_ppregstrength}), 
let us first
consider the piece of 
Eq.~(\ref{eq_llprime})
in square brackets.
If we decompose the incoming wave $\Phi(\vec{r})$ into partial waves,
\begin{eqnarray}
\Phi(\vec{r}) = \sum_{n_il_im_i} c_{n_il_im_i} Q_{n_il_i}(r) 
Y_{l_im_i}(\theta,\phi),
\end{eqnarray}
where the $c_{n_il_im_i}$ denote expansion coefficients and the
$Q_{n_il_i}$ radial basis functions,
the spherical harmonic $Y_{l0}$ in the integrand of $V_{ll'}$ 
acts as a projector or filter. After the integration 
over the angles, only those components of $\Phi(\vec{r})$
that have $l_i=l$ and $m_i=0$ survive.
The
operator $\frac{\partial^{2l+1}}{\partial r^{2l+1}} r^{l+1}$ 
in Eq.~(\ref{eq_llprime}) is designed 
to then first cure the $r^{-l-1}$ divergencies of the $Q_{n_il}$, which arise
in the $r \rightarrow 0$ limit, and to then second
``extract'' the coefficients of the regular part of the $Q_{n_il}(r)$
that go as $r^{l}$~\cite{huan57}. 
Alltogether, this shows that  
the square bracket in 
Eq.~(\ref{eq_llprime}) reduces to a constant when the $r \rightarrow 0$
limit is taken.
To understand the remaining pieces
of the pseudo-potential,
we multiply Eq.~(\ref{eq_llprime}) from the left with
$Q_{n_ol_o}^* Y^*_{l_om_o}$ and integrate over all space.
The spherical harmonic $Y_{l'0}$ in Eq.~(\ref{eq_llprime}) then ensures that 
the integral is only non-zero when
$l'=l_o$ and $m_o=0$.
When performing the radial integration,
the $\delta(r)/r^{l'}$ term ensures that the coefficients of the regular part  
of the $Q_{n_ol_o}$ that go as $r^{l_o}$ are being extracted
(note that the remaining $1/r^2$ term cancels the $r^2$ in
the volume element).

Alltogether,
the analysis outlined in the previous paragraph shows that the 
functional form of $V_{ll'}$ 
ensures that  
the divergencies of the radial parts of the incoming and outgoing
wave is cured in the $r \rightarrow 0$ limit 
and 
that the
$l$th component of the incoming wave is scattered into the
$l'$th partial wave.
The sum over all $l$ and $l'$ values in Eq.~(\ref{eq_ppreg})
guarantees that any state with quantum number $l$ can be coupled
to any state with quantum number $l'$,
provided the corresponding generalized scattering length $a_{ll'}$ is
non-zero.
We note that the regularized pseudo-potential given by Eqs.~(\ref{eq_ppreg})
through (\ref{eq_ppregstrength}) is only appropriate if the external confining
potential in Eq.~(\ref{eq_ham}) 
has spherical symmetry~\cite{idzi06}. Generalizations
of the above zero-range pseudo-potential, aimed at treating interacting dipoles
under elongated confinement, 
require the regularization scheme to be modified to additionally cure
divergencies of
cylindrically symmetric wave functions.
These extensions will be subject of future studies.

We now discuss the 
generalized scattering lengths $a_{ll'}$, which determine
the scattering strengths of the $V_{ll'}$.
The $a_{ll'}$ have units
of length and are defined through the K-matrix
elements $K_{lm}^{l'm'}$~\cite{newt},
\begin{eqnarray}
\label{eq_scatt}
a_{ll'} = \lim_{k \rightarrow 0} \frac{-K_{l0}^{l'0}(k)}{k}
\end{eqnarray}
for $m=0$.
The scattering lengths $a_{ll'}$ and
$a_{l'l}$ are identical 
because the K-matrix is symmetric.
In general, the scattering lengths $a_{ll'}$ have to be determined from the 
K-matrix elements for the ``true'' interaction potential, 
which contains the long-range dipolar 
and a short-ranged repulsive part, of two interacting dipoles.
As discussed further in Sec.~\ref{sec_ho},
an approach along these lines is used
to obtain the squares shown in Fig.~\ref{fig3}.

Alternatively, 
it has been shown that the K-matrix elements 
(except for $K_{00}^{00}$, see below) for realistic 
potentials,
such as for the
Rb-Rb potential in a strong electric field~\cite{yi00} 
or an OH-OH model potential~\cite{rone06},
are approximated with high accuracy by the 
K-matrix elements
for the dipolar potential only, calculated in the first Born approximation.
Applying the Born approximation to the 
second term on the right hand side
of Eq.~(\ref{eq_dipole}), we 
find for $m=0$ and $l=l'$ ($l \ge 1$)
\begin{eqnarray}
\label{eq_born1}
a_{ll}= -\frac{2D_*}{(2l-1)(2l+3)},
\end{eqnarray}
and for $m=0$ and $l=l' + 2$
\begin{eqnarray}
\label{eq_born2}
a_{l,l-2} = -\frac{D_*}{(2l-1) \sqrt{(2l+1)(2l-3)}}.
\end{eqnarray}
For $l'=2$ and $l=0$, 
e.g., Eq.~(\ref{eq_born2}) reduces to $a_{20}=-D_*/(3 \sqrt{5})$,
in agreement with Ref.~\cite{dere03}.
The scattering lengths $a_{l-2,l}$ are equal to
$a_{l,l-2}$, and 
all other generalized 
scattering lengths are zero. In Eqs.~(\ref{eq_born1}) and 
(\ref{eq_born2}), $D_*$ denotes the dipole length, 
$D_* = \mu d^2/\hbar^2$.
All non-zero scattering lengths $a_{ll'}$ are negative, 
depend on $l$ and $l'$, and are
directly proportional to $d^2$.
Furthermore, for fixed $D_*$, the absolute value of 
the non-zero $a_{ll'}$ decreases with increasing 
angular momentum quantum number $l$, indicating that
the coupling between different angular momentum
channels decreases with increasing $l$. However, this decrease is quite
slow and, in general, an accurate description of the
two-dipole system requires that the convergence with increasing 
$l_{max}$ be assessed carefully.

One can now show readily
that the K-matrix elements $K_{l0}^{l'0}$ of $V_{pp,reg}$,
calculated in the first
Born approximation, 
with $a_{ll'}$ given by Eqs.~(\ref{eq_born1}) and (\ref{eq_born2})
coincide with the K-matrix elements $K_{l0}^{l'0}$ of $V_{pp}$. This 
provides a simple check of the 
zero-range pseudo-potential construction and proofs
that the prefactors of $V_{ll'}$ are correct.
In turn, this suggests that the applicability regimes
of $V_{pp}$ and $V_{pp,reg}$ are comparable, if the 
generalized scattering lengths $a_{ll'}$ 
used to quantify the scattering strengths of $V_{ll'}$ 
are approximated by
Eqs.~(\ref{eq_born1}) and (\ref{eq_born2}). The applicability
regime of $V_{pp,reg}$ may, however, be larger than that of $V_{pp}$
if the full energy-dependent K-matrix of a realistic potential is used
instead.

To generalize the zero-range pseudo-potential defined
in Eqs.~(\ref{eq_ppreg}) through (\ref{eq_ppregstrength}) 
for projection
quantum numbers $m=0$ to any $m$, only a few changes have to be made.
In Eq.~(\ref{eq_llprime}), the spherical harmonics $Y_{l0}$
have to be replaced by $Y_{lm}$,
and
the generalized scattering lengths 
have to be defined through $\lim_{k \rightarrow 0} -K_{lm}^{l'm'}/k$.
Correspondingly, Eqs.~(\ref{eq_born1}) and (\ref{eq_born2})
become $m$-dependent.

\section{Two dipoles under external confinement}
\label{sec_ho}
Section~\ref{sec_hoA} derives the implicit eigenequation for two
dipoles interacting through the pseudo-potential under external harmonic 
confinement and Section~\ref{sec_hoB} analyzes the resulting eigen spectrum.

\subsection{Derivation of the eigenequation}
\label{sec_hoA}
To determine the eigen energies of two aligned dipoles
with $m=0$ under spherical harmonic confinement interacting through
the zero-range potential $V_{pp,reg}$, we 
expand the eigenfunctions $\Psi(\vec{r})$ in terms of
the 
orthonormal harmonic oscillator eigen functions $R_{n_il_i}Y_{l_i0}$,
\begin{eqnarray}
\label{eq_expansion}
\Psi(\vec{r}) = \sum_{n_il_i} c_{n_il_i} R_{n_il_i}(r) Y_{l_i0}(\theta,\phi).
\end{eqnarray}
The pseudo-potential $V_{pp,reg}$ enforces the proper boundary condition 
of $\Psi(\vec{r})$ at $r=0$, and thus determines the expansion
coefficients
$c_{n_il_i}$.
To introduce the key ideas we
first consider $s$-wave interacting particles~\cite{busc98}, for which
the pseudo-potential reduces to a single term, and then consider the
general case, in which the pseudo-potential contains infinitely
many terms.

Including only the term with $l$ and $l'=0$ in Eq.~(\ref{eq_ppreg}),
the Schr\"odinger equation becomes,
\begin{eqnarray}
\label{eq_swave}
\sum_{n_il_i} c_{n_il_i}
(E_{n_il_i} - E + V_{00}) 
R_{n_il_i}(r) Y_{l_i0}(\theta,\phi)
=0,
\end{eqnarray}
where the $E_{n_il_i}$ denote the eigenenergies of the non-interacting
harmonic oscillator,
\begin{eqnarray}
\label{eq_hoen}
E_{n_il_i}=\left( 2 n_i + l_i + \frac{3}{2} \right) \hbar \omega.
\end{eqnarray}
In what follows, it is convenient
to express the energy $E$ of the interacting system in
terms of a non-integer quantum number $\nu$,
\begin{eqnarray}
\label{eq_nu}
E = \left( 2 \nu + \frac{3}{2} \right) \hbar \omega.
\end{eqnarray}
Multiplying Eq.~(\ref{eq_swave}) from the left with 
$R^*_{n_ol_o}Y^*_{l_o0}$ with $l_o>0$ and integrating over all space,
we find that the $c_{n_il_i}$
with $l_i>0$ vanish. 
This can be understood readily by realizing that the
$s$-wave pseudo-potential $V_{00}$, as discussed in detail
in Sec.~\ref{sec_pp}, only couples states with $l=l'=0$.
To determine the expansion coefficients $c_{n_i0}$, we
multiply Eq.~(\ref{eq_swave}) from the left with 
$R^*_{n_o0}Y^*_{00}$ and integrate over all space.
This results in
\begin{eqnarray}
\label{eq_swave1}
c_{n_o0} (2 n_o - 2 \nu) \hbar \omega +
R_{n_o0}^*(0) g_{00}
B_0 =0,
\end{eqnarray}
where $B_0$ denotes the result of the square bracket in Eq.~(\ref{eq_llprime}),
\begin{eqnarray}
\label{eq_swave2}
B_0 = \left[ 
\frac{\partial}{\partial r} \left( r \sum_{n_i=0}^{\infty} 
c_{n_i0} R_{n_i0}(r) \right)
\right]_{r \rightarrow 0}.
\end{eqnarray}
Note that $B_0$ is constant and independent of $n_i$.
In Eq.~(\ref{eq_swave1}), the $r$-independent term $R_{n_o0}^*(0)$ arises
from
the radial integration over the $\delta$-function
of the pseudo-potential.
If we solve Eq.~(\ref{eq_swave1})
for $c_{n_o0}$ and
plug
the result into Eq.~(\ref{eq_swave2}),
the unknown constant $B_0$ cancels and we obtain an
implicit eigenequation
for $\nu$,
\begin{eqnarray}
1 = g_{00}
\left[ \frac{\partial}{\partial r} \left( r \sum_{n_i=0}^{\infty} 
\frac{R_{n_i0}^*(0) R_{n_i0}(r)}
{(2 \nu - 2 n_i) \hbar \omega} \right) \right]_{r \rightarrow 0}.
\end{eqnarray}
Using Eqs.~(\ref{eq_app1}) and (\ref{eq_app6}) 
from the Appendix to simplify the
term in square brackets,
we obtain the well-known 
implicit eigenequation for two
particles interacting through the $s$-wave pseudo-potential
under spherical harmonic confinement~\cite{busc98},
\begin{eqnarray}
\label{eq_swavefinal}
\frac{\Gamma \left( \frac{-E}{2 \hbar \omega}+\frac{1}{4} \right) }
{2 \Gamma \left( \frac{-E}{2 \hbar \omega} + \frac{3}{4} \right)} - 
\frac{a_{00}}{a_{ho}} =0.
\end{eqnarray}
Here, $a_{ho}$ denotes the harmonic oscillator length, 
$a_{ho}=\sqrt{\hbar/(\mu \omega)}$.

The derivation of the implicit eigenequation 
for two dipoles under external harmonic
confinement interacting through the pseudo-potential with infinitely
many terms proceeds analogously to that outlined above for the
$s$-wave system. The key difference is that each $V_{ll'}$ term
in Eq.~(\ref{eq_llprime}) with $l \ne l'$ 
couples states with 
different angular momenta,
resulting in a set of coupled equations for the expansion
coefficients $c_{n_il_i}$.
However, since $V_{pp,reg}$ for dipolar systems
couples only angular momentum states with $|l-l'| \le 2$
[see, e.g., the discussion at the beginning of
Sec.~\ref{sec_pp} and around Eqs.~(\ref{eq_born1}) and
(\ref{eq_born2})],
the coupled equations can, as we outline in the following, 
be solved analytically by including successively
more terms in $V_{pp,reg}$.

To start with, we plug the expansion given in Eq.~(\ref{eq_expansion})
into Eq.~(\ref{eq_se}), where the interaction potential
$V_{int}$ is now taken to be the pseudo-potential $V_{pp,reg}$ with
infinitely
many terms.
To obtain the general equation for the expansion coefficients $c_{n_il_i}$,
we multiply as before from the left with $R^*_{n_ol_o}Y^*_{l_o0}$
and integrate over all space,
\begin{eqnarray}
\label{eq_general1}
c_{n_ol_o} (2 n_o + l_o - 2\nu) \hbar \omega +
\left [\frac{R^*_{n_ol_o}(r)}{r^{l_o}} 
\right]_{r \rightarrow 0} \times \nonumber \\
\left[ 
g_{l_o-2,l_o} B_{l_o-2} +
g_{l_ol_o} B_{l_o}+ 
g_{l_o+2,l_o} B_{l_o+2} \right]=0.
\end{eqnarray}
Here, the $B_{l_o-2}$, $B_{l_o}$ and $B_{l_o+2}$ denote 
constants that are independent of $n_i$,
\begin{eqnarray}
\label{eq_general2}
B_{l_o} = \left[ 
\frac{\partial^{2l_o+1}}{\partial r^{2l_o+1}} \left\{ r^{l_o+1} \left( 
\sum_{n_i=0}^{\infty} c_{n_il_o} R_{n_il_o}(r)
\right) \right\}
\right]_{r \rightarrow 0}.
\end{eqnarray}
The three terms in the square bracket in the second line
of Eq.~(\ref{eq_general1}) arise because the $V_{l'-2,l'}$,
$V_{l'l'}$ and $V_{l'+2,l'}$ terms in the pseudo-potential $V_{pp,reg}$
couple the state $R^*_{n_ol_o}Y^*_{l_o0}$, for $l'=l_o$, with three components
of the expansion for $\Psi$, Eq.~(\ref{eq_expansion}).
Importantly, the constants $B_{l_o-2}$, $B_{l_o}$ and 
$B_{l_o+2}$, defined in Eq.~(\ref{eq_general2}), 
depend on the quantum numbers $l_o-2$, $l_o$ and $l_o+2$, 
respectively, which implies that Eq.~(\ref{eq_general1})
defines a set of infinitely many coupled equations that determine,
together with Eq.~(\ref{eq_general2}), the expansion coefficients
$c_{n_il_i}$.
Notice that Eqs.~(\ref{eq_general1}) and (\ref{eq_general2}) coincide
with Eqs.~(\ref{eq_swave1}) and (\ref{eq_swave2})
if we set $l_o=0$ and $g_{ll'}=0$ if $l$ or $l' > 0$.

We now illustrate how 
Eqs.~(\ref{eq_general1}) and (\ref{eq_general2})
can be solved for identical bosons, i.e., in the case where
$l$ and $l'$ are even (the derivation for
identical fermions proceeds analogously). 
Our strategy is to solve these equations by
including successively more terms in the coupled equations,
or equivalently, in the pseudo-potential. 
As discussed above,
if $a_{00}$ is the only non-zero scattering length, the
eigenenergies are given by Eq.~(\ref{eq_swavefinal}).
Next, we also allow for non-zero $a_{20}$, $a_{02}$ and $a_{22}$,
i.e., we consider $l$ and $l' \le 2$ in Eq.~(\ref{eq_ppreg}).
In this case, the coefficients $c_{n_i0}$ and $c_{n_i2}$ are non-zero
and coupled, but all $c_{n_il_i}$ with $l_i>2$ are zero.
Using the expressions for $B_0$ and $B_2$
given in Eq.~(\ref{eq_general2}),
we decouple the equations.
Finally, using Eqs.~(\ref{eq_app1}) and
(\ref{eq_app6}) from the Appendix,
the eigenequation
can be compactly written as
\begin{eqnarray}
\label{eq_uptotwo}
t_0+ \frac{q_2}{t_2}=0,
\end{eqnarray}
where 
\begin{eqnarray}
\label{eq_tl}
t_l = \frac{\Gamma( \frac{-E}{2 \hbar \omega} + \frac{1}{4} - \frac{l}{2})}
{2^{2l+1} \Gamma(\frac{-E}{2 \hbar \omega} +\frac{3}{4} + \frac{l}{2} )}
-
(-1)^l \frac{a_{ll}}{k^{2l} a_{ho}^{2l+1}},
\end{eqnarray}
and
\begin{eqnarray}
\label{eq_ql}
q_l = -\frac{a_{l-2,l}^2}{k^{4l-4}a_{ho}^{4l-2}}.
\end{eqnarray} 
Equation~(\ref{eq_uptotwo}) can be understood as follows.
If only $a_{00}$ is non-zero, it reduces to $t_0=0$, in
agreement with Eq.~(\ref{eq_swavefinal}).
If only $a_{00}$, $a_{02}$ and $a_{20}$ are non-zero,
Eq.~(\ref{eq_uptotwo}) remains valid if $a_{22}$ in $t_2$
is set to zero. This shows that 
the term $q_2$ and the first term on the right hand side of $t_2$
arise due to the coupling between states with
angular momenta $0$ and $2$. The second term of $t_2$, in contrast, arises
due to a non-zero $a_{22}$.
Finally, for non-zero $a_{00}$ and $a_{22}$ but vanishing 
$a_{20}$ and $a_{02}$, Eq.~(\ref{eq_uptotwo}) reduces to $t_0t_2=0$.
In this case, we recover the eigenequations $t_0=0$ for $s$-wave interacting
particles~\cite{busc98} and $t_2=0$ for $d$-wave interacting 
particles~\cite{stoc04}.

We now consider $l$ and $l'$ values with up
to $l_{max}=4$ in Eq.~(\ref{eq_ppreg}), i.e., we additionally allow for non-zero
$a_{24}$, $a_{42}$ and $a_{44}$, and discuss how the solution changes
compared to the $l_{max}=2$ case. 
The equation for the expansion coefficients 
$c_{n_i0}$ remains unchanged while that 
for $c_{n_i2}$ is modified. Furthermore,
the expansion coefficients $c_{n_i4}$ are no longer zero.
Consequently, we have three coupled equations,
which can be decoupled, resulting in the
following implicit eigenequation,
$t_0+q_2/(t_2+q_4/t_4)=0$. 
In analogy to the $l_{max}=2$ case, the $q_4$ term and the first part 
on the right hand side of
the 
$t_4$ term arise due to the ``off-diagonal'' scattering lengths $a_{24}$ and
$a_{42}$, and the second term of $t_4$ arises due to the ``diagonal''
scattering length $a_{44}$.

Next, let us assume that 
we have found the implicit eigenequation for the case
where we include terms in Eq.~(\ref{eq_ppreg})
with $l$ and $l'$ up to $l_{max}-2$.
If we now include terms 
with $l$ and $l'$ up to $l_{max}$, only the
equations for the expansion coefficients $c_{n_o l_o}$ 
with $l_o = l_{max}-2$ and $l_{max}$ change; those for the 
expansion coefficients $c_{n_ol_o}$ with $l_o \le l_{max}-4$ remain unchanged.
This allows the $l_{max}/2+1$ 
coupled equations for the expansion coefficients to be decoupled analytically
using the results already determined for the case where $l$ and $l'$ 
go up to $l_{max}-2$.
Following this procedure, 
we find the following implicit eigenequation
\begin{eqnarray}
\label{eq_eigen}
T_{l_{max}}=0,
\end{eqnarray}
where $T_{l_{max}}$ itself can be written as a continued fraction. 
For identical bosons we find,
\begin{eqnarray}
\label{eq_tlmax}
T_{l_{max}} = t_0 +
\frac{q_2}{t_2 + \frac{q_4}{t_4 + \cdots +\frac{q_{l_{max}}}{t_{l_{max}}}}}.
\end{eqnarray}
Taking $l_{max} \rightarrow \infty$ gives the eigenequation
for two identical bosons under spherical harmonic confinement 
interacting through $V_{pp,reg}$ with infinitely many terms.
For two identical fermions, Eqs.~(\ref{eq_tl})
through
(\ref{eq_tlmax})
remain valid if the subscripts $0,2,\cdots$ in Eq.~(\ref{eq_tlmax})
are
replaced by $1,3,\cdots$.

The derived eigenequation reproduces the eigenenergies in the known limits.
For the non-interacting case (all $a_{ll'}=0$), the eigenenergies
coincide with the eigenenergies of the harmonic oscillator, 
i.e.,
$E_{nl}= ( 2n + l + 3/2 ) \hbar \omega$,
where $n=0,1,2,\cdots$ and $l=0,2,4,\cdots$
(in the case of identical bosons) and 
$l=1,3,\cdots$ (in the case of identical fermions).
The $k$th levels, with energy $(2k+ 3/2) \hbar \omega$ for bosons
and $(2k+5/2) \hbar \omega$ for fermions, has a degeneracy
of $k+1$, $k=0,1,\cdots$.
Non-vanishing $a_{ll'}$ lead to a splitting
of degenerate energy levels but leave the number of energy
levels unchanged. 
If $a_{ll}$ is the only non-zero scattering length, the eigenequation
reduces to that obtained for spherically symmetric
pseudo-potentials with partial wave $l$~\cite{stoc04}.

\subsection{Analysis of the energy spectrum}
\label{sec_hoB}
This section analyses the implicit eigenequation, Eq.~(\ref{eq_eigen}),
derived in the previous section for the zero-range pseudo-potential
for $m=0$ and compares the resulting 
energy spectrum with that obtained for a shape-dependent model potential.
The implicit eigenequation, Eq.~(\ref{eq_eigen}), 
can be solved readily numerically by
finding its roots in different energy regions.
The solutions of the Schr\"odinger equation
for the shape-dependent model potential are otained by 
expanding the eigenfunctions on a B-spline basis.

\begin{figure}
      \centering
      \includegraphics[angle=270,width=8cm]{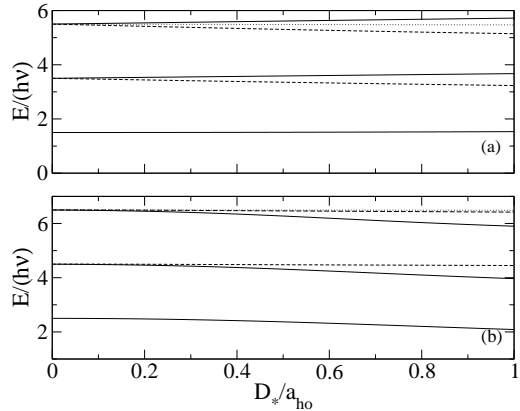}
\caption{
Relative eigenenergies $E$ for (a) two identical bosonic 
dipoles and (b) two identical
fermionic dipoles interacting
through $V_{pp,reg}$ [using $a_{00}=0$ in (a)]
under spherical harmonic confinement 
as a function of $D_*/a_{ho}$.
The line style indicates the predominant character of the
corresponding eigenstates. In (a), a solid line refers to $l \approx 0$,
a dashed line to $l \approx 2$, and a dotted line to $l \approx 4$;
in (b), a solid line refers to $l \approx 1$,
a dashed line to $l \approx 3$, and a dotted line to 
$l \approx 5$.
}
\label{fig1}
\end{figure}

Lines in Figs.~\ref{fig1}(a) and (b)
show the eigenenergies obtained by solving 
Eq.~(\ref{eq_eigen}) for two identical bosons and two identical
fermions, respectively,
interacting through $V_{ps,reg}(\vec{r})$ 
under external spherically symmetric harmonic
confinement 
as a function of the dipole length $D_*$.
In both panels, we assume that the interaction between the two dipoles 
is purely dipolar, i.e., in Fig.~\ref{fig1}(a) 
we set $a_{00}=0$.
The other scattering lengths $a_{ll'}$ are approximated by
Eqs.~(\ref{eq_born1}) and (\ref{eq_born2}).
Interestingly, for identical bosons, the lowest gas-like level,
which starts at $E=1.5 \hbar \omega$ for $D_*=0$, increases with 
increasing $D_*$. For identical fermions, in contrast,
the lowest gas-like state decreases with increasing $D_*$.

In addition to obtaining the eigenenergies themselves, the 
pseudo-potential treatment allows the spectrum to be classified
in terms of angular momentum quantum numbers. To this end,
we solve the implicit eigenequation, Eq.~(\ref{eq_eigen}),
for increasing $l_{max}$, and monitor how the energy levels
shift as additional angular momenta are
included in
$V_{pp,reg}$. 
Since a level with approximate quantum number $l$ changes only little as 
larger angular momentum values are included in the pseudo-potential,
this analysis reveals the predominant
character of each energy level. 
In Fig.~\ref{fig1}(a), the eigenfunctions of energies shown 
by solid, dashed and dotted lines 
have predominantly $l=0$, 2 and 4 character, respectively.
In Fig.~\ref{fig1}(b), the eigenfunctions of energies shown 
by solid, dashed and dotted lines 
have predominantly $l=1$, 3 and 5 character, respectively.
We find that the lowest excitation frequency between states 
with predominantly 
$l=0$ [$l=1$] character, increases [decreases] for identical
bosons [fermions] with increasing $D_*$. 
These predictions can be verified directly experimentally.

To assess the accuracy of the developed 
zero-range pseudo-potential treatment,
we consider two interacting bosons with non-vanishing
$s$-wave scattering length $a_{00}$. 
We imagine that the dipole moment of two identical
polarized bosonic polar molecules is tuned by an external electric
field. 
As the dipole moment $d$ is tuned,
the $s$-wave scattering length $a_{00}$, which
depends on the short-range and the
long-range physics of the ``true'' interaction potential, 
changes.
To model this situation,
we solve the two-body Schr\"odinger equation,
Eq.~(\ref{eq_se}), numerically for a shape-dependent
model potential with hardcore radius $b$ and long-range
dipolar tail. In this case, $V_{int}$ is given by 
\begin{eqnarray}
\label{eq_model}
V_{model}(\vec{r}) = 
	\left\{ \begin{array}{ll}
        d^2\frac{1 - 3 \cos^2 \theta} {r^3} & \mbox{if $r \ge b$}\\
        \infty & \mbox{if $r < b$} \end{array} \right. .
\end{eqnarray}
For $d=0$, the $s$-wave scattering length $a_{00}$
for $V_{model}$ is given by $b$. As the dipole length $D_*$ increases,
$a_{00}$ goes through zero, 
and becomes negative. Just when the two-body potential
supports a new bound state, $a_{00}$ goes through a resonance
and becomes large and positive.
As $D_*$ increases further, $a_{00}$ decreases. This resonance structure 
repeats itself with increasing $D_*$ (see Fig.~1 of Ref.~\cite{bort06};
note, however, that the lengths $a_{ho}$ and $D_*$ defined
throughout the present work differ from those defined in Ref.~\cite{bort06}).

\begin{figure}
      \centering
      \includegraphics[angle=270,width=8cm]{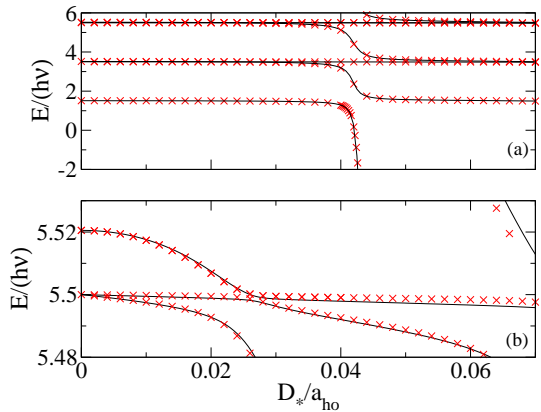}
\caption{
Panel~(a) shows the relative energies $E$ for two aligned identical bosonic
dipoles
under external spherical harmonic confinement 
as a function of $D_*/a_{ho}$.
Solid
lines show the numerically determined energies obtained using $V_{model}$
with $b=0.0097 a_{ho}$.
Crosses show the energies obtained using $V_{pp,reg}$ 
with essentially infinitely many
terms, and $a_{00}$ calculated for $V_{model}$. 
Panel~(b) shows a blow-up of the energy
region around $E \approx 5.5 \hbar \omega$.
Note that the horizontal axis in (a) and (b) are identical.
}
\label{fig2}
\end{figure}

For the model potential $V_{model}$,
$a_{00}$ depends on the ratio between the short-range and long-range
length scales, i.e., on $b/D_*$.
To compare the pseudo-potential energies and the energies for the model
potential, we fix $b$ and calculate $a_{00}$ for each $D_*$ 
considered. The dipole-dependent
$s$-wave scattering length is then used in the
zero-range pseudo-potential $V_{pp,reg}$.
The other scattering lengths are, as before,
approximated by the expressions given in Eqs.~(\ref{eq_born1}) 
and (\ref{eq_born2}).
Solid lines in Fig.~\ref{fig2}(a) and (b) show the eigenenergies
obtained
for $V_{model}$ as a function of $D_*$.
Crosses show the eigenenergies obtained for 
$V_{pp,reg}$ using a value of $l_{max}$ that results in converged
eigenenergies. 
The overview spectrum shown in Fig.~\ref{fig2}(a)
shows that one of the energy levels dives down to negative energies
close to that $D_*$ value at which the two-body potential $V_{model}$ 
supports a new bound state.
The blow-up, Fig.~\ref{fig2}(b), around
$E \approx 5.5 \hbar \omega$
shows excellent agreement between the energies obtained 
using $V_{pp,reg}$ (crosses) and those obtained using 
$V_{model}$ (solid lines); the maximum deviation for the energy range
shown is 0.05~\%.

As before, we can assign approximate quantum numbers to each energy level.
At $D_* \ll a_{ho}$, 
the three energy levels around $E \approx 5.5 \hbar \omega$ 
have,
from bottom to top, approximate quantum numbers $l=2$, 4 and 0. After
two closely spaced
avoided crossings around $D_* \approx 0.025 a_{ho}$, the assignment
changes to $l=0$, 2 and 4 (again, from bottom to top).
If the maximum angular momentum $l_{max}$ of the pseudo-potential is set  
to 2, the energy level with approximate quantum
number $l=4$ would be absent
entirely.
This illustrates that a complete and accurate description of the energy 
spectrum requires the use of a zero-range pseudo-potential with 
infinitely many terms.
The energy of
a state with approximate quantum number $l$ requires $l_{max}$ 
to be at least $l$ for the correct degeneracy be obtained and
at least $l+2$ for a quantitative description.

The sequence of avoided crossings at $D_* \approx 0.025 a_{ho}$
suggests an interesting experiment. Assume that the system is initially,
at small electric field (i.e., small $D_*/a_{ho}$), 
prepared in the excited state with angular momentum $l \approx 0$ and
$E \approx 5.52 \hbar \omega$. The electric field is then slowly swept
across the first broad avoided crossing at $D_* \approx 0.019 a_{ho}$ 
to transfer the population from the
state with $l \approx 0$ to the state with $l \approx 2$. 
We then suggest to sweep quickly across the second 
narrower avoided crossing at $D_* \approx 0.028 a_{ho}$
(the ramp speed must be chosen so minimize population
transfer from the state with $l \approx 2$ to the state with $l \approx 4$). 
As in the case of $s$-wave scattering only~\cite{dunn04},
the time-dependent field sequence has to be optimized to 
obtain maximal population transfer.
The proposed scheme promises to provide an efficient means for
the transfer of population between states with different angular momenta
and for quantum state engineering.

\begin{figure}
      \centering
      \includegraphics[angle=270,width=8cm]{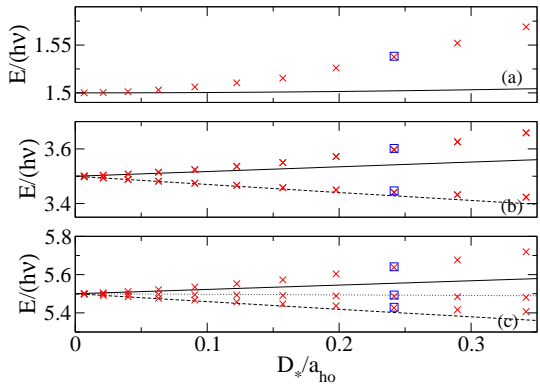}
\caption{
Crosses show the
relative eigenenergies $E$ as a function of $D_*/a_{ho}$ 
for two identical bosons with $a_{00}=0$ interacting through
$V_{model}$ with $b = 0.0031 a_{ho}$ 
in three different energy regions.
Lines show $E$ 
for two identical bosons with $a_{00}=0$ interacting through
$V_{ps,reg}$ with
$a_{ll'}$ given by Eq.~(\protect\ref{eq_born1}) and 
(\protect\ref{eq_born2}).
As in Fig.~\protect\ref{fig1}(a) solid, dashed and dotted
lines show the energies of levels 
characterized by approximate quantum numbers $l \approx 0$, 2 and 4.
The agreement between the crosses and the lines is good at small
$D_*/a_{ho}$ but less good at larger $D_*/a_{ho}$. 
Squares show the eigenenergies obtained for the energy-dependent
pseudo-potential at $D_*=0.242a_{ho}$;
the agreement between the squares and the crosses is excellent,
illustrating
that usage of the energy-dependent
K-matrix greatly enhances the applicability regime of $V_{pp,reg}$.
}
\label{fig3}
\end{figure}

Figure~\ref{fig2} illustrates that the pseudo-potential
treatment reproduces the eigenenergies of the shape-dependent
model potential $V_{model}$. 
To further assess the validity of the pseudo-potential treatment,
we now consider two interacting bosonic
dipoles for which the dipolar interaction is
dominant, i.e., we consider $a_{00}=0$.
For $V_{model}$ with $b = 0.0031 a_{ho}$, we determine a set
of $D_*$ values at
which $a_{00}=0$.
Note that 
the number of bound states with predominantly
$s$-wave character increases by one for each successively
larger $D_*$. 
Crosses in Figs.~\ref{fig3}(a)-(c) show the eigenenergies
for $V_{model}$ with $a_{00}=0$ as a function of $D_*$ in the energy ranges
around $1.5$, $3.5$ and $5.5 \hbar \omega$. 
For comparison, lines show
the eigenenergies obtained for the regularized pseudo-potential
with $a_{00}=0$.
As in Fig.~\ref{fig1}, the linestyle indicates the predominant character
of the energy levels
(solid line: $l \approx 0$;
dashed line: $l \approx 2$;
and
dotted line: $l \approx 4$).
The agreement between the energies obtained for
the pseudo-potential with $a_{ll'}$ given by
Eqs.~(\ref{eq_born1}) and (\ref{eq_born2}) and for the model
potential for small $D_*$ is very good, thus
validating the applicability of the pseudo-potential treatment. 
The agreement becomes less good, however, as 
$D_*$ increases. This can be explained readily by realizing
that the dipole length $D_*$ approaches the harmonic
oscillator length $a_{ho}$. 

In general, the description of confined particles 
interacting 
through zero-range
pseudo-potentials is justified if the characteristic lengths of the
two-body potential are smaller than the characteristic length of the
confining potential. For example, in the case of 
$s$-wave interactions only, 
the van der Waals length has to be smaller than the oscillator 
length~\cite{blum02,bold02}.
The model potential $V_{model}$ is characterized by
a short-range length scale, the hardcore radius $b$, and
the dipole length $D_*$; in Fig.~\ref{fig3}, it is the relatively
large value of $D_*/a_{ho}$ that leads, eventually,
to a break-down of the pseudo-potential treatment. 
As in the case of spherical
interactions, the break-down can be pushed to larger $D_*$
values by introducing energy-dependent generalized 
scattering lengths $a_{ll'}(k)$, defined through
$-K_{l0}^{l'0}(k)/k$ for $m=0$, and 
by then solving the eigenequation, Eq.~(\ref{eq_eigen}),
self-consistently~\cite{blum02,bold02}.

\begin{figure}
      \centering
      \includegraphics[angle=270,width=8cm]{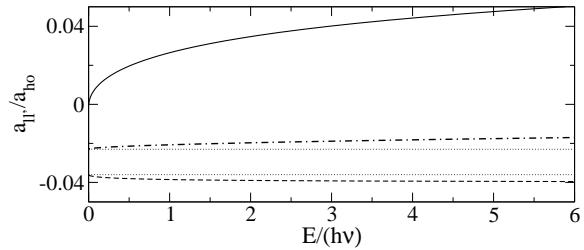}
\vspace*{-0.6in}
\caption{Energy-dependent scattering lengths $a_{00}(k)$
(solid line), $a_{20}(k)$ (dashed line) and $a_{22}(k)$ (dash-dotted line)
for the model potential $V_{model}$ with $D_*=78.9b$ 
as a function of the relative energy $E$. In oscillator units,
$V_{model}$ is characterized by
$b=0.0031 a_{ho}$ and $D_*=0.242 a_{ho}$. For comparison,
horizontal dotted lines show the energy-independent
scattering lengths $a_{22}$, Eq.~(\protect\ref{eq_born1}), and 
$a_{20}$, Eq.~(\protect\ref{eq_born2}), calculated in the 
first Born approximation.
}
\label{fig4}
\end{figure}

Figure~\ref{fig4} shows three selected scattering lengths $a_{ll'}(k)$
for the model potential $V_{model}$
with $D_*=78.9b$ as a function of energy.
This two-body potential supports eight bound states
with projection quantum number $m=0$, which have
predominantly $s$-wave character.
Both energy and length in Fig.~\ref{fig4} are expressed in oscillator units
to allow for direct comparison with the data shown in Fig.~\ref{fig3}.
The scattering length $a_{00}(k)$, shown by a solid line
in Fig.~\ref{fig4}, is zero at zero energy and increases with increasing
energy.
Both $a_{20}(k)$ (dashed line)  and $a_{22}(k)$ (dash-dotted line)
are negative. Their zero-energy values coincide with those calculated
in the Born approximation (horizontal dotted lines).

Using these energy-dependent $a_{ll'}(k)$ to parametrize
the strengths of the pseudo-potential 
and solving the eigenequation, Eq.~(\ref{eq_eigen}), self-consistently, 
we 
obtain the squares in Fig.~\ref{fig3}.
The energies for $V_{pp,reg}$ 
with {\em{energy-dependent}} $a_{ll'}$ (squares)
are in much better agreement with the energies obtained for the 
model potential (crosses) than the energies obtained using the
{\em{energy-independent}} $a_{ll'}$ to parametrize the pseudo-potential
(lines).
This suggests that the applicability regime of the
regularized zero-range pseudo-potential
can be extended significantly by introducing energy-dependent
scattering lengths.
Since the proper treatment of resonant interactions within the
regularized zero-range pseudo-potential requires that 
the energy-dependence of the generalized scattering lengths 
be included, future work will address this issue in more depth.

\section{Summary}
\label{sec_conclusion}
This paper applies a zero-range pseudo-potential treatment to
describe two interacting dipoles under external spherically
harmonic confinement.
Section~\ref{sec_pp}
introduces the regularized zero-range
pseudo-potential $V_{pp,reg}$ used in this work, which was first
proposed by Derevianko~\cite{dere03,dere05}.
Particular emphasis
is put on developing a simple interpretation of the
individual pieces of the pseudo-potential. Furthermore, we clearly
establish the connection between 
$V_{pp,reg}$ and the pseudo-potential $V_{pp}$, 
which is typically employed 
within a mean-field framework. We argue that the applicability regime of these
two pseudo-potentials is comparable if the scattering strengths of 
$V_{pp,reg}$, calculated in the 
first Born approximation, are chosen so as to reproduce those of $V_{pp}$.

We then use the regularized 
zero-range pseudo-potential to derive an implicit eigen
equation for two dipoles under external confinement, a system which
can be realized experimentally with the aid of optical lattices. 
In deriving the implicit eigenequation, we again put emphasis 
on a detailed understanding of how the solution arises, thus
developing a greater understanding of the underlying physics.
The
implicit eigenequation can be solved straightforwardly,
and allows for a direct classification scheme
of the resulting eigenspectrum.
By additionally calculating the eigen energies for two
dipoles interacting through a finite range model potential
numerically, we assess the applicability of the developed
zero-range pseudo-potential treatment. We find good agreement
between the two sets of eigenenergies for small $D_*$ and
quantify the deviations as $D_*$ increases.
Finally, we show that the validity regime of $V_{pp,reg}$
can be extended by parametrizing the scattering strengths
of $V_{pp,reg}$ 
in terms of the energy-dependent K-matrix calculated for a realistic model 
potential.
This may prove useful also when describing resonantly interacting dipoles.

At first sight it may seem counterintuitive 
to replace the {\em{long-range}} dipolar interaction 
by a {\em{zero-range}} pseudo-potential. However, 
if the length scales of the interaction potential,
i.e., the van der Waals length scale characterizing the
short-range part and the dipole length characterizing the long-range
part of the potential,
are smaller than the characteristic length of the trap $a_{ho}$,
this approach is justified 
since the
zero-range pseudo-potential is designed to reproduce the 
K-matrix elements of the ``true'' interaction potential. 
This is particularly true if the pseudo-potential is taken to
contain infinitely many terms, as done in this work.

In summary,
this paper determines the eigenenergies of two interacting dipoles
with projection quantum number $m=0$.
The applied zero-range pseudo-potential treatment is validated by comparing the
resulting eigenenergies with those
obtained numerically for a shape-dependent model potential.
The analysis presented sheds further light on the intricate
properties of angle-dependent scattering processes and their
description through a regularized 
zero-range pseudo-potential with infinitely many
terms. The calculated energy spectrum may aid on-going
experiments on dipolar Bose and Fermi gases.

Acknowledgements:
KK and DB acknowledge support by the NSF through grant PHY-0555316
and JLB by the DOE.

\section{Appendix}
In this Appendix, we evaluate the following infinite sum,
\begin{eqnarray}
\label{eq_app1}
C_l = 
\left[ \frac{\partial^{2l+1}}{\partial r^{2l+1}} \left\{ r^{l+1}
\sum_{n=0}^{\infty}
\frac{
\left[\frac{R^*_{nl}(r)}{r^l}\right]_{r\rightarrow0}
R_{nl}(r)}
{2 \left(\nu-n-\frac{l}{2} \right) \hbar \omega}  \right\}
\right]_{r \rightarrow 0}.
\end{eqnarray}
Writing the radial harmonic oscillator functions $R_{nl}(r)$ in terms
of the Laguerre polynomials $L_n^{(l+1/2)}$,
\begin{eqnarray}
\label{eq_app2}
R_{nl}(r)=
\sqrt{\frac{2^{l+2}}{(2l+1)!!\pi^{1/2}L_n^{(l+1/2)}(0) a_{ho}^{3}}}
\times \nonumber \\
\exp \left(-\frac{r^2}{2a_{ho}^2} \right) 
\left( \frac{r}{a_{ho}} \right)^l 
L_n^{(l+1/2)}(r^2/a_{ho}^2),
\end{eqnarray}
we find
\begin{eqnarray}
\label{eq_app3}
\left[\frac{R_{nl}(r)}{r^l}\right]_{r\rightarrow0}=
\sqrt{\frac{2^{l+2}L_n^{(l+1/2)}(0)}{(2l+1)!!\pi^{1/2} a_{ho}^{2l+3}}}.
\end{eqnarray}
Using Eqs.~(\ref{eq_app2}) and (\ref{eq_app3}), the $C_l$
can be rewritten as
\begin{eqnarray}
\label{eq_app4}
C_l=
\frac{2^{l+1}}{(2l+1)!!\pi^{1/2} a_{ho}^{2l+3}} \times \nonumber \\
\left[\frac{\partial^{2l+1}}{\partial r^{2l+1}} \left(
\exp\left(\frac{-r^2}{2a_{ho}^2}\right) 
r^{2l+1}\sum_{n=0}^{\infty}{\frac{L_n^{(l+1/2)}\left((\frac{r}{a_{ho}})^2 \right)} 
{\left(\nu -n-\frac{l}{2} \right) \hbar \omega}} \right) 
\right]_{r\rightarrow0}.
\end{eqnarray}
We evaluate the infinite sum in Eq.~(\ref{eq_app4}) 
using the properties of the generating function~\cite{abranote1},
\begin{eqnarray}
\label{eq_app5}
\sum_{n=0}^{\infty}{\frac{L_n^{(l+1/2)}((r/a_{ho})^2)}{\nu -n-\frac{l}{2}}}=
\nonumber \\
-\Gamma(-\nu + l/2)U(-\nu + l/2, l+3/2, (r/a_{ho})^2).
\end{eqnarray}
Using Eq.~(\ref{eq_app5}) together with the small $r$
behavior of the hypergeometric function $U$~\cite{abranote2},
the expression for the $C_l$ reduces to
\begin{eqnarray}
\label{eq_app6}
C_l =
\frac{(-1)^l 2^{2l+2}(2l)!!}{(2l+1)!!}
\frac{\Gamma \left(-\nu + \frac{l}{2} \right)}
{\Gamma \left(-\nu-\frac{l+1}{2} \right)} \frac{1}{\hbar \omega a_{ho}^{2l+3}}.
\end{eqnarray}

%\bibliography{lit}
%\bibliographystyle{prsty}
\end{document}